\documentclass[11pt]{article}

\usepackage{graphics,graphicx,fullpage,natbib, multirow}
\usepackage{amsmath,amssymb,verbatim,epsfig}
\usepackage[colorlinks=true,linkcolor=black!50!blue,citecolor=black!20!blue,urlcolor=black!10!blue]{hyperref}
\usepackage[dvipsnames,usenames]{color}
\usepackage{graphicx}
\usepackage{subcaption}
\usepackage[normalem]{ulem}
\usepackage{enumitem}
\usepackage{soul}
\usepackage{setspace} 
\usepackage{appendix}
\usepackage{xcolor}

\newtheorem{theorem}{Theorem}

\newtheorem{assumption}{Assumption}

\newtheorem{remark}{Remark}

\newtheorem{subassumption}{Assumption}[assumption]

\newtheorem{defin}{Definition}

\newcommand{\bI}{ {\bf I} }
\newcommand{\bX}{ {\bf X} }
\newcommand{\bY}{ {\bf Y} }
\newcommand{\bZ}{ {\bf Z} }

\newcommand{\bU}{ {\bf U} }

\newcommand{\bV}{ {\bf V} }

\newcommand{\tr}{ \mbox{tr}}

\newcommand{\bA}{ {\bf A} }

\newcommand{\bL}{ {\bf L} }
\newcommand{\bD}{ {\bf D} }
\newcommand{\bE}{ {\bf E} }
\newcommand{\bS}{ {\bf S} }

\newcommand{\bSigma}{ {\mbox{\boldmath $ \Sigma $} }}

\newcommand{\bPhi}{ {\mbox{\boldmath $ \Phi $} }}

\newcommand{\bzero}{ {\bf 0} }

\newcommand{\bB}{ {\bf B} }

\newcommand{\Td}{T_{\text{dev}}}

\newcommand{\bXd}{{\bf X}_{\text{dev}}}
\newcommand{\bYd}{{\bf Y}_{\text{dev}}}

\title{\textbf{Boosting multi-view association testing via  devariation}}

\date{}
{\singlespacing
\author{Ruyi Pan$^{1,2}$, Yinqiu He$^{3,*}$, Jun Young Park$^{1,4,}$\thanks{Co-corresponding authors: \url{yinqiu.he@wisc.edu}, \url{junjy.park@utoronto.ca}}}
    \date{
    {\small $^1$ {Department of Statistical Sciences, University of Toronto, Toronto, ON, Canada}}\\
    {\small $^2$ {Centre for Addiction and Mental Health, Toronto, ON, Canada}}\\
    {\small $^3$ {Department of Statistics, University of Wisconsin-Madison, Madison, WI, USA}}\\
    {\small $^4$ {Department of Psychology, University of Toronto, Toronto, ON, Canada}}
}
}

\begin{document}

\maketitle

{\singlespacing
\begin{abstract} \noindent
  Understanding the interplay between high-dimensional data from different views is essential in biomedical research, particularly in fields such as genomics, neuroimaging and biobank-scale studies involving high-dimensional features. Existing statistical tests for the association between two random vectors often do not fully capture dependencies between views due to limitations in modeling within-view dependencies, particularly in high-dimensional data without clear dependency patterns, which can lead to a potential loss of statistical power. In this work, we propose a novel approach termed \textit{devariation} which is considered a simple yet effective preprocessing method to address the limitations by adopting a penalized low-rank factor model to flexibly capture within-view dependencies. Theoretical analysis of asymptotic power  shows that devariation increases statistical power, especially when within-view correlations impact signal-to-noise ratios, while maintaining robustness in scenarios without strong internal correlations. Simulation studies demonstrate devariation's superior performance over existing methods in various scenarios. We further validate devariation in multimodal neuroimaging data from the UK Biobank study, examining the associations between imaging-derived phenotypes (IDPs)  from functional, structural, and diffusion {magnetic resonance imaging (MRI)}. \vspace{3mm}
  
  
\end{abstract}
}

   

\section{Introduction}\label{sec:intro}

Improving the statistical power of association tests between two random vectors is a fundamental problem in statistics.
Such tests have provided valuable insights into complex systems with biomedical applications such as genetics, genomics, neuroscience, etc. For instance, in genome-wide association studies (GWASs), a set-based association testing between genetic variants (single nucleotide polymorphisms) and multiple phenotypes identified multiple genetic loci beyond what univariate testing provided.
Similarly, brain-wide association studies (BWASs), which focus on the relationship between brain imaging features and behavioral or neurocognitive features, have revealed how brain activity is linked to cognitive functions or mental health conditions. Recent large-scale studies  identified extensive phenotypic and genetic associations between heart and brain MRI traits derived from the UK Biobank study, illustrating the need for powerful methods to detect complex cross-view dependencies in high-dimensional data \citep{zhao2023heart}.

A wide array of statistical testing methods has been developed to test for associations between two random vectors, extending beyond linear relationships to capture  nonlinear or conditional associations. The RV coefficient proposed by \cite{robert1976unifying} was an early instance of generalizing the notion of
correlation to the multivariate context. Significance testing based on the RV coefficient, termed RV test in this paper, can be used to test the linear association between two random vectors. Canonical Correlation Analysis (CCA) is a popular statistical method used to explore the relationships between two sets of variables by finding pairs of linear combinations that are maximally correlated, which can be used to test associations between two sets of variables \citep{bartlett1941statistical}.
Moving beyond linear associations, \cite{szekely2007measuring} proposed distance correlation (dCor) by measuring the distance between the joint characteristic function of the two random vectors and the product of their marginal characteristic functions. This is closely related to the Hilbert–Schmidt Independence Criterion (HSIC) \citep{gretton2005measuring}. 
In addition to comparing characteristic functions, several approaches have emerged that focus on comparing density functions or cumulative distribution functions \citep{berrett2019nonparametric, heller2013consistent}.

In the existing literature, various methods have been proposed to account for dependencies in data to facilitate inference.  
One important improvement involves the use of generalized estimating equations (GEE), which directly model the dependencies of the response variable through a working correlation matrix
\citep{wang2013gee,zhang2014testing,park2022clean,pan2024spatial}. Another approach to account for these relationships is to use kernels, as \cite{dutta2019multi} applied kernels to {genotypes} and phenotypes {separately}. This approach is closely related to distance-based or kernel-based tests \citep{zhan2017fast,minas2013distance}, which can also be viewed as distance-based multivariate regression analysis \citep{zapala2006multivariate,han2010powerful,shi2023distance}. However, both working correlation matrices and kernels come with the inherent risk of misspecification, and the performance of these tests heavily depends on the chosen working correlation matrices or kernels. As discussed by \cite{wang2003working}, misspecifying the working correlation matrix would cause a substantial loss in statistical power. Although field-specific kernels have been developed,  as discussed in \cite{broadaway2016statistical}, their specialized nature often limits their generalizability and applicability across different contexts. Choosing appropriate kernels or working correlation structures requires deeper knowledge of the underlying data structure, and modeling the covariance structure remains a critical challenge.

\begin{figure}[h]
    \centering
    \includegraphics[scale=0.5]{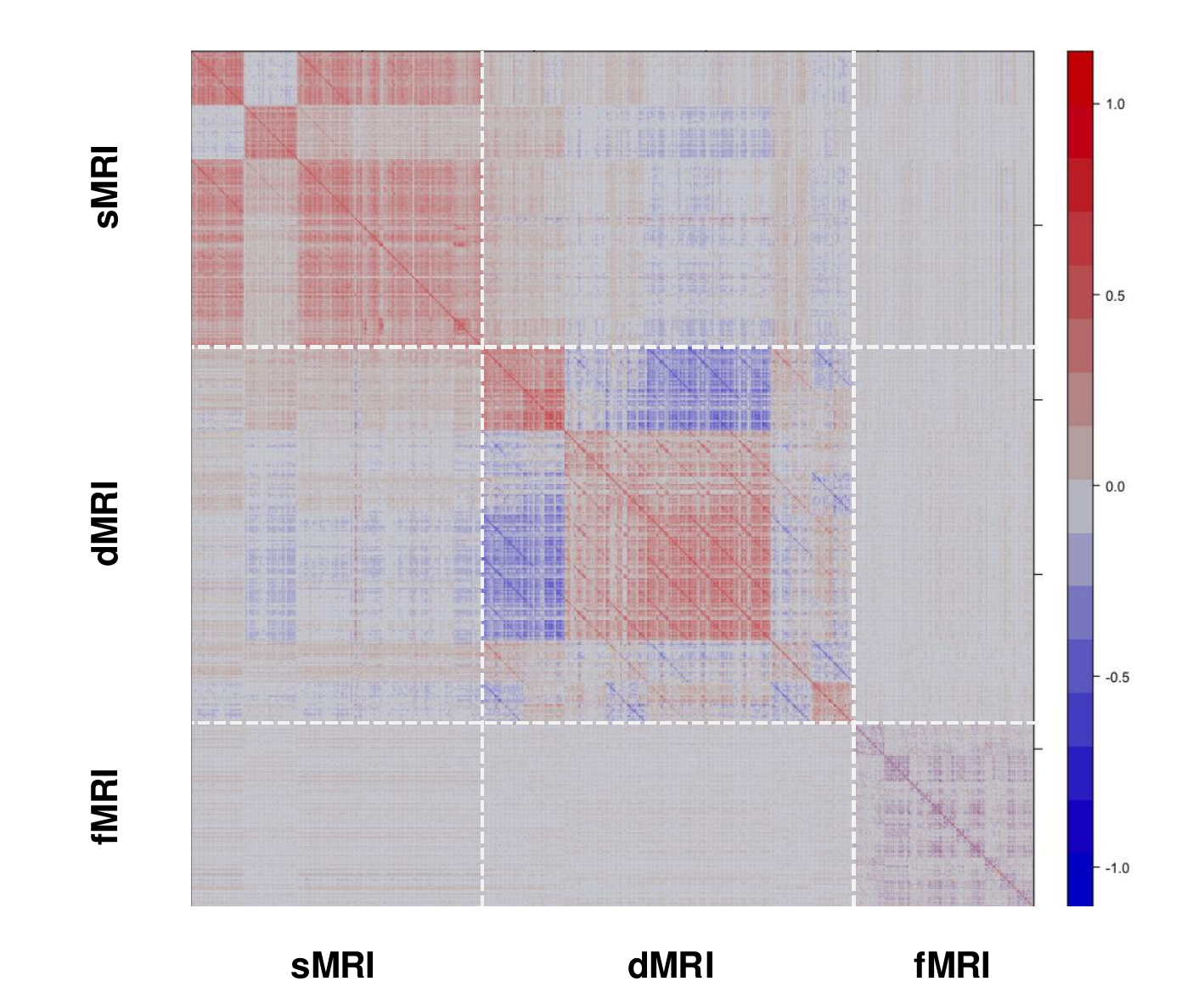}
    \caption{ Correlation matrix generated from three-view imaging-derived phenotypes (IDPs) from the ‘population-level’ neuroimaging data from the UK
Biobank study with $n=39,587$ subjects. There are 339 features in the sMRI view, including white surface area, thickness, and volume,  grey matter volume, and subcortical volume.  dMRI contains 432 skeleton measurements covering metrics like Fractional Anisotropy (FA), Intracellular Volume Fraction (ICVF), Isotropic Volume Fraction (ISOVF), the eigenvalues of the diffusion tensor (L1, L2, L3), Mean Diffusivity (MD), Mode of Anisotropy (MA), and Orientation Dispersion (OD).  The fMRI view includes 210 resting-state functional connectivity (rsFC) features.} 
    \label{fig:corrplot_full}
\end{figure}

Existing work has shown that parametric modeling of each modality's covariance structure via a Gaussian process improves statistical power \citep{weinstein2022spatially}. However, this challenge becomes even more pronounced in high-dimensional data without clear structured covariance, where multiple sets of features or measurements are collected from different views. For example, we consider imaging-derived phenotypes (IDPs) from the population-level brain imaging data from the UK Biobank study with $n=39,587$ subjects, including structural MRI (sMRI), diffusion MRI (dMRI), and functional MRI (fMRI), which characterize the brain's anatomy,  microstructure, and function, respectively. As shown in Figure \ref{fig:corrplot_full}, the correlation structure within each imaging view is unclear and varies significantly across views, making it challenging to specify a suitable working correlation structure to efficiently model the dependencies within each imaging view. Although using sample covariances is a common approach to model dependencies \citep{dutta2019multi}, we note that even the sample covariance matrix is inconsistent in high-dimensional settings, when the number of features is of the same order as the sample size.

This paper proposes a flexible approach to accommodate dependencies within each view to improve the statistical power of global association testing in high dimensions. We use a penalized low-rank factor model in a data-driven way to efficiently capture dependencies, which is particularly useful in high-dimensional  data with nontrivial dependence structures. Recently, low-rank factor models have shown promise in modeling  multi-view data, providing meaningful and effective representations of complex structures. For example,  Joint and Individual Variation Explained (JIVE) \citep{lock2013joint} and related methods such as
angle-based JIVE \citep{feng2018angle}, D-CCA \citep{shu2020d}, and BIDIFAC/BIDIFAC+ \citep{park2020integrative, lock2022bidimensional} have become increasingly prominent tools in  integrative analysis of genomics and imaging. These methods are adept at disentangling joint variation shared commonly among different data views and individual variations distinctive to each data view. Leveraging these models to construct a more powerful inference procedure has been largely unexplored, and we aim to address this gap in this paper.

The proposed method, devariation, consists of two steps. 
Given $n$ independent samples of two random vectors represented by matrices $\bX \in \mathbb{R}^{n\times p}$ and $\bY\in \mathbb{R}^{n\times q}$, we separately shrink the singular values of $\bX$ and $\bY$ to a certain level guided by  random matrix theory. The `devariation-adjusted' $\bX$ and $\bY$, obtained after  shrinkage, are then used as inputs to existing methods. Although devariation utilizes the  singular value soft-thresholding, our contribution goes beyond what existing work on matrix completion addressed for a single matrix \citep{mazumder2010spectral}. Specifically,  devariation-adjusted data are often regarded as noise that is typically discarded in the matrix completion literature \citep{donoho1995noising,cai2010singular}. Although it is counter-intuitive, we offer a new model-based perspective showing that the devariation-adjusted data, when used as intermediate quantities for association testing, can carry meaningful within-view dependence and enhance power. To our knowledge, this perspective  has not been explicitly formulated or systematically studied in the existing literature. Under this regime, we show that the proposed method has the following advantages: 
\begin{enumerate}[nolistsep]
    \item \textit{Flexibility.}
     Devariation is suitable for high-dimensional and multi-view datasets without the need to explicitly model and estimate parametric structures of correlations.

     \item \textit{Simplicity.} 
     Devariation can be considered a simple preprocessing step that separately shrinks two data matrices, making it computationally efficient and easy to implement.

    \item \textit{Adaptiveness to within-view dependencies.}
     In this paper, we focus on applying devariation to the RV test, given its close connections to  the popular tests used in biomedical or genetic data,  built upon the linear model framework. We show that it achieves high power when the within-view dependence is well captured by low-rank structures. It also incurs little adverse effect when within-view dependencies are weak.
\end{enumerate}

The rest of the paper is structured as follows. In Section \ref{sec:method}, we describe how to construct a test statistic with devariation and provide its model-based interpretation. In Section \ref{sec:theory}, we provide theoretical guarantees of devariation applied to the RV test under different scenarios. In Section \ref{sec:simulation}, we conduct extensive simulations to evaluate the Type I error rate and statistical power of the proposed method, and compare its performance to other tests. We conduct data-driven simulations using IDPs from the UK Biobank to evaluate the robustness of devariation in Section \ref{sec:real_data}. Finally, 
a discussion of the proposed method is   provided in Section \ref{sec:discussion}. Additional details and proofs are deferred to the Supplementary Material.

We summarize the notation used below. For two sequences of real numbers $\{f_n\}$ and $\{h_n\}$: $f_n=O(h_n)$ means that $|f_n|\leqslant c|h_n|$ for a constant $c > 0$; $f_n \ll h_n$ means $\lim_{n\rightarrow\infty}f_n/h_n=0$;
$f_n \gg h_n$ means $\lim_{n\rightarrow\infty}h_n/f_n=0$. The matrix normal distribution is denoted by $\mathcal{MN}$.
Let $\bI_p$ denote the $p\times p$  identity matrix, and $\bzero_{p\times q}$ denote $p\times q$ the zero matrix. For matrices,  $\lVert \cdot \rVert_{F}$ and $\lVert \cdot \rVert_{*}$ denote the Frobenius norm and the nuclear norm, respectively;  tr(·) denotes the trace operator. $\bPhi(\cdot)$ is the cumulative distribution function of $\mathcal{N}(0,1)$.

\section{Method}\label{sec:method}

\subsection{Setup}
Let $\bX\in\mathbb{R}^{n\times p}$  and $\bY\in\mathbb{R}^{n\times q}$ denote the data from view 1 and view 2, respectively, where $n$ is the number of samples, $p$ 
and $q$ 
are the numbers of features from views 1 and 2, respectively. Without loss of generality, we assume {that each feature (column) of} $\bX$ and $\bY$  has been standardized. 

\subsection{RV coefficient and related methods}\label{subsec:rv_gee}
In the following, we review the RV coefficient \citep{robert1976unifying}, based on which to illustrate our proposed approach of devariation. The RV coefficient has the following form:
\begin{align}\label{eq:rv_coeff}
    \text{RV}(\bX,\bY) = \frac{\tr(\bX\bX^{\top}\bY\bY^{\top})}{\sqrt{\tr(\bX\bX^{\top}\bX\bX^{\top})\cdot \tr(\bY\bY^{\top}\bY\bY^{\top})}}.
\end{align}
Note that
    $\text{RV}(\bX,\bY)$ is proportional to the sum of squared pairwise correlations between all features of $\bX$ and $\bY$, thereby quantifying their overall linear association.
From Equation \eqref{eq:rv_coeff}, the denominator of the  RV coefficient is permutation-invariant, and the numerator induces power for testing the association between $\bX$ and $\bY$. Therefore, we take the numerator of Equation \eqref{eq:rv_coeff} as the test statistic in this paper, denoted by
\begin{align}\label{eq:T_RV}
    T_{\text{RV}}(\bX,\bY) = \tr(\bX\bX^{\top}\bY\bY^{\top}).
\end{align}

It is well-known that  Equation \eqref{eq:T_RV} can be viewed as a specific instance of score test statistic for testing $H_0:\sigma_B^2=0$ in a linear random effect model when $\bX$ is fixed:\begin{align}\label{model:random_effect1}
     \bY=\bX\bB+\bE_Y, \quad \bB \sim \mathcal{MN}(\bzero_{p\times q}, \bI_p, \sigma_B^2\bI_q), \quad  \bE_Y \sim \mathcal{MN}(\bzero_{n\times q}, \bI_n, \bSigma_{Y}),
\end{align}
under the assumption of $\bSigma_{Y}=\tau_Y^2\bI_q$. Under the model assumptions, score test for $H_0$ is known to be the locally most powerful test for $\sigma_B^2$ close to 0 \citep{lin1997variance}. Alternatively, one can also consider modeling $\bX$  {when} $\bY$ is fixed:
\begin{align}\label{model:random_effect2}
     \bX=\bY\bA+\bE_X, \quad \bA \sim \mathcal{MN}(\bzero_{q\times p}, \bI_q, \sigma_A^2\bI_p), \quad  \bE_X \sim \mathcal{MN}(\bzero_{n\times p}, \bI_n, \bSigma_{X}),
\end{align}
with an assumption of $\bSigma_X=\tau_X^2\bI_p$, which yields the same test statistic in Equation \eqref{eq:T_RV} up to a scaling constant for testing $H_0: \sigma_A^2=0$. 

Model formulations in Equations \eqref{model:random_effect1} and \eqref{model:random_effect2} imply that $T_{\text{RV}}$ may be underpowered when dependencies exist among features in $\bX$ or $\bY$ (i.e., $\bSigma_{Y}\neq \tau_Y^2\bI_q$ or $\bSigma_{X}\neq \tau_X^2\bI_p$). This limitation is particularly relevant in real-world situations where within-view dependence is common and strong (e.g., spatial dependence in surface-based neuroimaging data). Alternative methods, such as Generalized Estimating Equations (GEE)-based score tests \citep{wang2013gee}, consider specifying a working covariance structure to model $\bSigma_X$ or $\bSigma_Y$, such as exchangeable, auto-regressive, or spatial Gaussian processes, leading to modified test statistics $T_{\text{RV}}(\bX, \bY\widetilde{\bSigma}_{Y}^{-1})$ or $T_{\text{RV}}(\bX\widetilde{\bSigma}_{X}^{-1}, \bY)$ under Equations  \eqref{model:random_effect1} and \eqref{model:random_effect2}, respectively, where  $\widetilde{\bSigma}_{Y}$ and  $\widetilde{\bSigma}_{X}$ are estimators of covariance matrices $\bSigma_X$ and $\bSigma_Y$ under $H_0$ (i.e., no linear association between $\bX$ and $\bY$).
The effectiveness of such approaches relies on selecting an appropriate correlation structure, which often requires external information.

One possible way to account for randomness in both $\bX$ and $\bY$ in the test statistic {is}
\begin{align}\label{eq:rv_score}
          T_{\text{RV}}(\bX\widetilde{\bSigma}_{X}^{-1}, \bY\widetilde{\bSigma}_{Y}^{-1}).
 \end{align}
This statistic falls into a broad class of kernel- or distance-based  tests \citep{zhan2017fast,minas2013distance}. 
For example, applying the linear kernel described in \cite{zhan2017fast} to both  $\bX$ and $\bY$ is equivalent to $T_{\text{RV}}(\bX,\bY)$. Other options are possible, such as the weighted linear kernel and the Gaussian kernel.
However, the performance of these methods depends heavily on the choice of kernel or distance metric, and choosing an appropriate one  can be challenging.

\subsection{Proposed method}
\subsubsection{Devariation}
In this section, we propose a new method called devariation, where the devariation-adjusted data $\bXd$ and $\bYd$ are defined as \begin{align}\label{eq:devariation}
\bXd=\bX-S_{\lambda_X}(\bX),~~\bYd=\bY-S_{\lambda_Y}(\bY).
\end{align} The operator $S_{\lambda_Y}(\bY)$  is defined via the  singular value soft-thresholding procedure as 
\begin{equation}\label{eq:soft_SVD}
    S_{\lambda_Y}(\bY)=\bU_Y\bD_{\lambda_Y}\bV_{Y}^T \quad\text{with} \quad \bD_{\lambda_Y}=\text{diag}[(d_1-\lambda_Y)_{+},\ldots,(d_r-\lambda_Y)_{+}], 
\end{equation}
where $\bU_Y\bD_{Y}\bV_{Y}^{\top}$ is the  singular value decomposition (SVD) of $\bY$ with $\bD_{Y}=\text{diag}[d_1,\ldots,d_r]$, and $(d_j-\lambda_Y)_{+}=\max(d_j-\lambda_Y,0)$ for $j=1,\dots,r$. As shown in \cite{mazumder2010spectral},  it is the solution to the optimization problem
\begin{align}\label{eq:soft_estimator}
   \underset{\bZ: \text{rank}(\bZ) \leqslant r}{\arg\min}\lVert \bY- \bZ\rVert_{F}^2 + 2\lambda_Y\lVert \bZ\rVert_{*}.
\end{align} 

With the devariation-adjusted $\bX$ and $\bY$, the test statistic of the devariation RV is 
\begin{align}\label{eq:rv_devariation}
    \Td = T_{\text{RV}}\left(\bXd,\bYd\right).
\end{align}
Since the test statistic is computed by replacing the $\bX$ and $\bY$ with the devariation-adjusted data ($\bXd$ and $\bYd$), which are obtained without jointly modeling the relationship between $\bX$ and $\bY$, existing approaches can still be used to compute a $p$-value. These include the method of \citet{josse2008testing} based on a Pearson type III approximation and the permutation-based approach of \citet{heo1998permutation}, which is implemented in the {\texttt{FactoMineR}} R package. Especially, when using permutation, it is not necessary to soft-threshold the singular values every time but is sufficient to permute rows of $\bX_{\text{dev}}$, which makes computational cost the same as RV test given $\bX_{\text{dev}}$  and  $\bY_{\text{dev}}$.

\vspace{-10pt}

\paragraph{Methodological  novelty.} 
We highlight that our focus is on the impact of devariation on  power in association testing, which substantially differs from existing studies on using the singular value soft-thresholding estimator itself for   matrix completion.  Moreover, compared to methods based on principal component regression or the existing tests that preserve a few top principal components of $\bY$ to be used for association testing \citep{dutta2019multi}, our method takes the `non-signal' components  $\bX-S_{\lambda_X}(\bX)$ and $\bY-S_{\lambda_Y}(\bY)$ where top singular values are shrunk to an appropriate level but not completely to 0. We show a model-based interpretation of the devariation RV test in Section \ref{sec:dev_interpretation} and investigate its asymptotic power in Section \ref{sec:theory}.

\subsubsection{Model-based interpretation of devariation RV }\label{sec:dev_interpretation}

In this section, we describe how $T_{\text{dev}}$ in Equation \eqref{eq:rv_devariation} is connected to Equation \eqref{eq:rv_score} using a latent variable formulation to model $\bX$ and $\bY$ simultaneously. We consider the probabilistic viewpoint of JIVE \citep{lock2013joint}  for its utility and empirical support in  multi-view  data. 
The JIVE model is formulated as \begin{align}\label{eq:low_rank_model}
    &\bX_{n\times p} = \bS_J\bL_{J_X}^{\top} + \bS_{I_X}\bL_{I_X}^{\top} + \bE_X,\nonumber\\
    &\bY_{n\times q} = \underbrace{\bS_J\bL_{J_Y}^{\top}}_{\text{Joint}} + \underbrace{\bS_{I_Y}\bL_{I_Y}^{\top}}_{\text{Individual}} + \underbrace{\bE_Y}_{\text{Noise}},
\end{align}
where 
\begin{itemize}[nolistsep]
       \item 
       The fixed loading matrices $\bL_{J_X}$, $\bL_{I_X}$ (for $\bX$) and $\bL_{J_Y}$, $\bL_{I_Y}$ (for $\bY$) characterize the joint and individual structures with respective ranks $r_J$, $r_{I_X}$, and $r_{I_Y}$.  For identifiability, we assume $\bL_{J_X}^{\top}\bL_{I_X} = \bzero_{r_J\times r_{I_X}}$,  $\bL_{J_X}^{\top}\bL_{J_X}=\bI_{r_J}$, and $\bL_{I_X}^\top\bL_{I_X}=\bI_{r_{I_X}}$.  Similar constraints hold for $\bL_{J_Y}$ and $\bL_{I_Y}$. 
  \item 
  The random effect matrices include shared scores $\bS_J$,  and individual scores $\bS_{I_X}$ and $\bS_{I_Y}$. $\bE_X$ and $\bE_Y$ are white noise matrices, where entries of each are independently distributed with mean zero and variances $\sigma_J^2, \sigma_{I_X}^2, \sigma_{I_Y}^2, \tau_X^2$  and $\tau_Y^2$, respectively.
\end{itemize}

Specifically, when  $p$ and $q$ are fixed, and the entries of random matrices are normally distributed, $\Td$ is naturally interpreted through model \eqref{eq:low_rank_model}. From our probabilistic model specification, the population covariance matrix is 
\begin{align}\label{eq:cov_low_rank}\begin{pmatrix}
        \sigma_J^2\bL_{J_X}\bL_{J_X}^\top+ \sigma_{I_X}^2\bL_{I_X}\bL_{I_X}^\top+\tau_X^2\bI_p & \sigma_J^2\bL_{J_X}\bL_{J_Y}^\top\\
        \sigma_J^2\bL_{J_Y}\bL_{J_X}^\top & \sigma_J^2\bL_{J_Y}\bL_{J_Y}^\top+ \sigma_{I_Y}^2\bL_{I_Y}\bL_{I_Y}^\top+\tau_Y^2\bI_q
    \end{pmatrix},
\end{align}
and the association between $\bX$ and $\bY$ are induced only by the shared score $\bS_J$ in the joint structure. Therefore, testing the existence of an association between $\bX$ and $\bY$ is equivalent to testing 
\begin{align*}
   H_0:\sigma_J^2=0~~\text{and}~~ H_1:\sigma_J^2>0.
\end{align*}
Note that covariances of $\bX$ and $\bY$ take the form of spiked covariance under both $H_0$ and $H_1$. Also, under $H_0$, we have
\begin{align}\label{eq:low_rank_model_H0}
    \bX=\bS_{I_X}\bL_{I_X}^{\top}+\bE_X,\quad \bY=\bS_{I_Y}\bL_{I_Y}^{\top}+\bE_Y,
\end{align}  
still allowing for parameterizing nontrivial within-view dependence of both $\bX$ and $\bY$. 
Under the current model, we note that Equation \eqref{eq:rv_score} is reformulated as 
\begin{align}\label{eq:rv_cond_expectation}
 T_{\text{RV}}(\bX\widetilde{\bSigma}_{X}^{-1}, \bY\widetilde{\bSigma}_{Y}^{-1}) 
 &\propto  T_{\text{RV}}(\bX\widetilde{\bSigma}_{X}^{-1}\tilde\tau_{X}^2, \bY\widetilde{\bSigma}_{Y}^{-1}\tilde\tau_{Y}^2)\notag\\
 &=T_{\text{RV}}(\widehat{\mathbb{E}}_{H_0}(\bE_X|\bX, \bL_{I_X}), \widehat{\mathbb{E}}_{H_0}(\bE_Y|\bY,\bL_{I_Y}))\notag\\
&= T_{\text{RV}}(\bX-\widehat{\mathbb{E}}_{H_0}(\bS_{I_X}\bL_{I_X}^\top|\bX, \bL_{I_X}), \bY-\widehat{\mathbb{E}}_{H_0}(\bS_{I_Y}\bL_{I_Y}^\top|\bY,\bL_{I_Y})),
\end{align}
where $\tilde\tau_{X}^2$ and $\tilde\tau_{Y}^2$ are the estimates of $\tau_X^2$ and $\tau_Y^2$ under $H_0$, implying that, with $\bL_{I_X}$ and $\bL_{I_Y}$ given,  it is sufficient to obtain the $\widehat{\mathbb{E}}_{H_0}(\bS_{I_X}\bL_{I_X}^\top|\bX,\bL_{I_X})$ and $\widehat{\mathbb{E}}_{H_0}(\bS_{I_Y}\bL_{I_Y}^\top|\bX,\bL_{I_Y})$ to construct the test statistic. One can also observe that Equation \eqref{eq:rv_cond_expectation} becomes the original test statistic in Equation \eqref{eq:T_RV} when within-view dependencies are not considered (i.e., $\bX=\bE_X$ and $\bY=\bE_Y$ in Equation \eqref{eq:low_rank_model_H0} without any individual
structure).

In practice, $\bL_{I_X}$ and $\bL_{I_Y}$ {as well as their ranks} are  unknown and need to be estimated. Principal component analysis (PCA) is one possible and popular approach to obtain $\bL_{I_X}$ and $\bL_{I_Y}$, but there are  potential over/underfitting issues. Therefore, we adopt the Bayesian framework and assign a Gaussian prior {(e.g., a ridge penalty)} to $\bL_{I_X}$ and $\bL_{I_Y}$ by {$\bL_{I_X} \sim \mathcal{MN}(\bzero,\bI_{\min\{n,p\}},\theta_{I_X}^2\bI_p)$ and} $\bL_{I_Y} \sim \mathcal{MN}(\bzero,\bI_{\min\{n,q\}},\theta_{I_Y}^2\bI_q)$. Taking $\bY$ as an illustrative example, we note that the logarithm  of the posterior distribution can be expressed as
\begin{align*}
    \log p(\bS_{I_Y},\bL_{I_Y}| \bY, \tau_Y^2, \sigma_{I_Y}^2) & = -\frac{1}{2\tau_Y^2}\lVert \bY- \bS_{I_Y}\bL_{I_Y}^{\top}\rVert_{F}^2-\frac{1}{2\sigma_{I_Y}^2}
   \lVert \bS_{I_Y}\rVert_{F}^{2} -\frac{1}{2\theta_{I_Y}^2}\lVert \bL_{I_Y}\rVert_{F}^{2}+\text{constant}.
\end{align*}
 Finding the maximum a posteriori (MAP) estimator of  $\bS_{I_Y}$ and $\bL_{I_Y}$, with fixed parameters $\tau_Y^2$ , $\sigma_{I_Y}^2$ and $\theta_{I_Y}^2$, is equivalent to 
\begin{align}\label{eq:objetive_min}
     \underset{{ \lbrace\bS_{I_Y}},{\bL_{I_Y}} \rbrace}{\arg\min} \lVert \bY- \bS_{I_Y}\bL_{I_Y}^{\top}\rVert_{F}^2 + \lambda_Y^{S}\lVert \bS_{I_Y}\rVert_{F}^{2} +\lambda_Y^{L}\lVert \bL_{I_Y}\rVert_{F}^{2} 
\end{align} 
where $\lambda_Y^{S} = \tau_Y^2/\sigma_{I_Y}^2$ and $\lambda_Y^{L} = \tau_Y^2/\theta_{I_Y}^2$.
Since $\min_{\bA=\bS_{I_Y}\bL_{I_Y}^\top}\lambda_Y^{S}||\bS_{I_Y}||_F^2+ \lambda_Y^{L}||\bL_{I_Y}||_F^2=2\sqrt{\lambda_Y^S\lambda_Y^L}||\bA||_*$,  $\widehat \bS_{I_Y} \widehat \bL_{I_Y}^{\top}$ obtained from  Equation \eqref{eq:objetive_min} reduces to $S_{\lambda_Y}(\bY)$ following Equation \eqref{eq:soft_estimator} with $\lambda_Y=\sqrt{\lambda_Y^S\lambda_Y^L}$. Therefore, we can obtain $S_{\lambda_Y}(\bY)$ which leads to  $\bYd= \bY-\widehat\bS_{I_Y} \widehat \bL_{I_Y}^{\top} =\bY- S_{\lambda_Y}(\bY)$. Applying it to $\bX$, one can obtain
    $\bXd= \bX-\widehat\bS_{I_X} \widehat \bL_{I_X}^{\top} =\bX - S_{\lambda_X}(\bX)$.
Then, the test statistic becomes $ T_{\text{RV}}(\bXd, \bYd)$, as given in Equation \eqref{eq:rv_devariation}, where the MAP estimator $S_{\lambda_Y}(\bY)$ replaces $\widehat{\mathbb{E}}_{H_0}(\bS_{I_Y}\bL_{I_Y}^{\top}|\bY,\bL_{I_Y})$ in Equation \eqref{eq:rv_cond_expectation}.

\subsubsection{Selecting  $\lambda_X$ and $\lambda_Y$}\label{sec:thresholds}
The tuning parameters  play a crucial role in determining the threshold of the singular values of $\bX$ and $\bY$, hence affecting statistical power. Taking $\bY$ as an example, choosing $\lambda_Y$ to efficiently obtain $\widehat{\bS}_{I_Y}\widehat{\bL}_{I_Y}^{\top}$ requires a low-rank assumption on $\bS_{I_Y}\bL_{I_Y}^\top$. First, from the random matrix result, the largest singular value has a tight {probabilistic} upper bound $\tau_Y(\sqrt{n} + \sqrt{q})$ \citep{rudelson2010non}, under the assumption that the entries of noise $\bE_{Y}$ are independent sub-Gaussian with variance $\tau_Y^2$.
 With the low-rankness assumption considered in the probabilistic JIVE model, 
 we choose this probabilistic upper bound as the threshold ($\lambda_Y = \tau_Y(\sqrt{n} + \sqrt{q}$)), and use  Equation \eqref{eq:soft_SVD} to estimate the low-rank matrix $\bS_{I_Y}\bL_{I_Y}^{\top}$.  This threshold has shown utility in the low-rank matrix reconstruction methods \citep{shabalin2013reconstruction,park2020integrative}, and it aligns with our goal of recovering $\mathbb{E}_{H_0}(\bE_Y|\bY)$ and $\mathbb{E}_{H_0}(\bE_X|\bX)$ in Section \ref{sec:dev_interpretation}. Since the proposed $\lambda_X$ and $\lambda_Y$  depend on the dimensionality of $\bX$ and $\bY$ and they determine the exact level to which the singular values of $\bX$ and $\bY$ are shrunk, we study the asymptotic power of the proposed method in Section \ref{sec:theory}.

 In practice, $\tau_X$ and $\tau_Y$  also need to be estimated. Given that  $\bS_J\bL_{J_Y}^{\top} + \bS_{I_Y}\bL_{I_Y}^{\top}$  is low-rank, it reduces to  the problem of estimating the standard deviation of a white noise matrix perturbed by low-rank signal, which has been widely studied. Therefore, we propose to use a robust estimator based on the median of empirical singular values and the median of Marchenko–Pastur distribution \citep{gavish2017optimal} as a default in this paper. There are other options such as \cite{shabalin2013reconstruction} which provides an approach to estimate
by minimizing the Kolmogorov-Smirnov distance between the empirical and theoretical singular value distribution.

\subsubsection{Remarks}
\begin{remark}
\normalfont
While our primary motivation for devariation is to test for a global association between two random vectors, it is also applicable when one of them is a scalar and only the within-view dependence of the other data view is modeled. In such a case, $T_{\text{RV}}$ is reduced to a score-based variance component test statistic in multiple  or multivariate linear regression, which are studied extensively in recent literature \citep{wu2011rare,dutta2019multi}.
\end{remark}

\begin{remark}
\normalfont
Under the Bayesian context, the Gaussian prior {on $\bL_{I_X}$ and $\bL_{I_Y}$} can be replaced with alternative priors/penalties, such as the Laplace prior for inducing sparsity.  While these options could be more useful than Gaussian prior when the within-view correlations are concentrated in a few features, we recognize the {practical and computational} challenges involved in selecting tuning parameters as well as ranks for $\bL_{I_X}$ and $\bL_{I_Y}$. 

\end{remark}

\section{Theory}\label{sec:theory}
This section demonstrates the asymptotic power gain    via  devariation and its adaptiveness based on the RV test in  a high-dimensional regime. 
For   concreteness, we establish asymptotic analysis under the random  latent factor model \eqref{eq:low_rank_model}, though we  expect the theoretical insights could be generalized to other scenarios as well. 

In particular,  we discuss three scenarios: (1) the individual structures dominate, (2) both the individual and joint structures are weak, and (3) the joint structure dominates. 
We first present a high level summary of our technical results, while details are given in Sections \ref{sec:Scenario2}--\ref{sec:Scenario3} below. 
The first scenario is expected to reflect the commonly observed patterns in biomedical data as demonstrated in Figure \ref{fig:corrplot_full}. In this case, we will show that the devariation effectively removes the variation of the individual structures and thus significantly boosts the   power of the association testing. 
In the second and third scenarios,   the individual structures have small variations, which are supposed to be less favorable to the devariation, as it may mistakenly remove more variations in the joint structure than in the individual structures. 
But interestingly, 
we   will show that the devariation strategy is adaptive to weak variations and maintains reasonably high power even in these adversarial scenarios. 




\paragraph*{Technical set-up} Before specifying three distinct scenarios, we  list the common assumptions used throughout this section. Notably, we do not need to   explicitly specify parametric distributions of the random components, further showing the flexibility of the framework.   

\begin{assumption}[Common assumptions]\label{assump:overall}
\normalfont
 Suppose that the model \eqref{eq:low_rank_model}  holds and 
  \begin{enumerate}[nolistsep]
        \item $\lim_{n\to \infty} p/n = c_X \in (0,1],~ \lim_{n\to \infty} q/n = c_Y \in (0,1]$ for some constants $c_X, c_Y$.
         \item  $r_J, r_{I_X}$,  and $r_{I_Y}$ are fixed positive integers.
        \item $\tau_X$ and $\tau_Y$ are fixed and positive constant.
        \item the entries of the scaled random matrix {($\bS_J/\sigma_J, \bS_{I_X}/\sigma_{I_X}, \bS_{I_Y}/\sigma_{I_Y},\bE_X/\tau_X$, and $\bE_Y/\tau_Y$)} are independent sub-Gaussian with unit variance, and finite eight moments.
    \end{enumerate}  
\end{assumption}

The first point reflects the reality of many modern multi-view datasets, where the number of features is comparable to the number of samples \citep{van2013wu}. In such settings, the dimensionality cannot be treated as fixed relative to the sample size, and high-dimensional asymptotic analysis becomes necessary.
The second point constrains the joint and individual structures to have sufficiently low rank, consistent with low-rank structures widely adopted in multi-view modeling \citep{lock2013joint}. Such finite-dimensional latent variation reflects the belief that a limited number of cross-view and within-view mechanisms drive the observed dependence patterns. The third assumption ensures that the noise variance remains at a fixed level, and the last assumption on the moments provides the regularity required to characterize the limiting behavior of the test statistics. These two assumptions together allow us to characterize the asymptotic power of the test statistics in terms of the relative magnitudes of joint and individual variation. In particular, with $\tau_X$ and $\tau_Y$ are fixed, varying $\sigma_J, \sigma_{I_X}$, or $\sigma_{I_Y}$ changes the corresponding ``signal-to-noise ratios''. 
Technically, we define three scenarios by varying $\sigma_J, \sigma_{I_X}$, and $\sigma_{I_Y}$ in Assumptions \ref{assump:S2}--\ref{assump:S3} below. 



Moreover, to streamline the discussions, we define a unified mapping to express the power function for $T_{\text{RV}}(\cdot,\cdot): \mathbb{R}^{n\times p}\times \mathbb{R}^{n\times q}\to \mathbb{R}$ as
\begin{align*}
    \beta_{\alpha}(\cdot,\cdot)=\mathbb{P}_{H_1}\left\{T_{\text{RV}}(\cdot,\cdot)> t_{1-\alpha}(\cdot,\cdot)\right\},
\end{align*}
where $(\cdot,\cdot)$ represents   the input two-view  data, and  $t_{1-\alpha}(\cdot,\cdot)$ denotes a corresponding critical value satisfying
\begin{align}\label{eq:typei}
\lim_{n\to\infty}\mathbb{P}_{H_0}\left\{T_{\text{RV}}(\cdot,\cdot)> t_{1-\alpha}(\cdot,\cdot)\right\}=\alpha, 
\end{align} i.e., asymptotically controlling the Type I error rate at significance level $\alpha$ under $H_0$.
Notably, the choice of the mapping $t_{1-\alpha}(\cdot,\cdot)$ satisfying \eqref{eq:typei} may not be unique, but  to fix ideas, we consider specific choices in our problem with 
\begin{align}
    t_{1-\alpha}(\bX,\bY)=q_{1-\alpha}\cdot(n\sigma_{I_X}^2\sigma_{I_Y}^2) \quad \text{ and }\quad 
     t_{1-\alpha}(\bXd,\bYd)=  z_{1-\alpha} \sqrt{\nu_0} + \mu_0, \label{eq:nu0mu0def} 
\end{align}
where $q_{1-\alpha}$ is the $(1-\alpha)$th level quantile of $\chi^2_{r_{I_X}r_{I_Y}}$, $z_{1-\alpha}$ is the $(1-\alpha)$th level quantile of $\mathcal{N}(0,1)$, $\nu_0$ is an approximation  for $\mathbb{V}_{H_0}(\Td)$, and $\mu_0$ is an approximation for $\mathbb{E}_{H_0}(\Td)$. We establish theoretical guarantees showing that these choices indeed satisfy \eqref{eq:typei} under our considered settings. See more details in Supplementary Material A.1.1 and  A.2.3 for the two critical values correspondingly.

\subsection{Scenario 1: dominant individual structures}\label{sec:Scenario2}

\begin{subassumption}\label{assump:S2} \normalfont
Suppose that  Assumption \ref{assump:overall} holds and
    \begin{enumerate}[nolistsep]
        \item[(i)] $\sigma_J \ll n^{1/2}$,  
        \item[(ii)] 
        $\sigma_{I_X}, \sigma_{I_Y} \gg n^{1/2}(1+\sigma_J^3)$. 
    \end{enumerate}
\end{subassumption}
 
 Under Assumption \ref{assump:S2}, $ \sigma_{I_X}, \sigma_{I_Y} \gg \sigma_J$, so the individual structures account for more data variation than the joint structure does, aligning with patterns shown in Figure \ref{fig:corrplot_full}. 
Theoretically, the cutoffs of ``strong'' and ``weak'' variations are defined relative to $n^{1/2}$ in the sense that $\sigma_J \ll n^{1/2}$ and $\sigma_{I_X}, \sigma_{I_Y} \gg n^{1/2}$. 
Using $n^{1/2}$ as a comparative baseline is reasonable because  the operator norms of the noise terms are of the order of $O(n^{1/2})$ by the random matrix theory \citep{rudelson2010non}. 
This is similarly used in Assumptions \ref{assump:S1}--\ref{assump:S3} below. 
Furthermore, Assumption \ref{assump:S2}   requires   $\sigma_{I_X}, \sigma_{I_Y} \gg n^{1/2}\sigma_J^3$ which is a purely technical condition to ensure that variations of the individual structures are substantially  larger than that of the joint strcuture when $\sigma_J\gg 1 $ and to facilitate our theoretical derivations. 


In Scenario 1, applying the devariation is expected to remove   prominent variations from the individual structures, which helps reduce variance of the test statistic and thus enhances the test power. To see this, we  rigorously characterize the asymptotic power of the RV test and the devariation RV test in Theorems \ref{thm:S2:power_original_data}--\ref{thm:S2:power_Tr}, respectively.  

\begin{theorem}
\label{thm:S2:power_original_data}\normalfont
Under Assumption \ref{assump:S2}, $\lim_{n\to\infty}\beta_{\alpha}(\bX,\bY)= \alpha$ under $H_1$.   
\end{theorem}

Theorem \ref{thm:S2:power_original_data} implies that the  RV test suffers from negligible power asymptotically when individual variation dominates the total variation. However, with devariation applied, the asymptotic power of  the RV test can be significantly improved, as shown in Theorem \ref{thm:S2:power_Tr}.

\begin{theorem}
 \label{thm:S2:power_Tr}
\normalfont
Under Assumption \ref{assump:S2}, when $\sigma_J$ is fixed, 
\begin{align}\label{eq:prop224_case1}
\lim_{n\to \infty}  \beta_{\alpha}(\bXd,\bYd)= 1-\Phi\left(z_{1-\alpha}-\frac{r_J(\sigma_J^4+c_Y\sigma_J^2\tau_Y^2+c_X\sigma_J^2\tau_X^2)}{\tau_X^2\tau_Y^2\sqrt{2c_Xc_Y+c_Xc_Y^2(\mu_{E_X,4}-1)+c_Yc_X^2(\mu_{E_Y,4}-1)}}\right).
\end{align}
Also, when $\sigma_J\to \infty$ as $n\to \infty$,  
\begin{align}\label{eq:prop224_case2}
   \lim_{n\to \infty}   \beta_{\alpha}(\bXd,\bYd) = 1,  
\end{align}
where $\mu_{E_X, 4}$ and $\mu_{E_Y, 4}$ are the fourth moments of the entries of
$\bE_X/\tau_X$ and $\bE_Y/\tau_Y$, respectively.
\end{theorem}

Since $\lim_{n\to\infty} \beta_{\alpha}(\bXd,\bYd) > \alpha$ as long as $\sigma_J$ does not degenerate to zero,
Theorems \ref{thm:S2:power_original_data} and \ref{thm:S2:power_Tr}   imply that  
$\lim_{n\to\infty} \beta_{\alpha}(\bXd,\bYd) > \lim_{n\to\infty} \beta_{\alpha}(\bX,\bY)$, 
demonstrating that devariation yields a strict improvement in asymptotic power. This scenario aligns with the motivation for devariation adjustment, as high-noise individualized components can be removed, thereby improving power. In less favorable scenarios,  we can show, interestingly, that devariation either preserves asymptotic power or results in only limited power loss.


\subsection{Scenario 2: weak joint and individual structures}\label{subsec:S1}

\begin{subassumption}
    \label{assump:S1} \normalfont

    Suppose that Assumption \ref{assump:overall}   holds and $\sigma_J, \sigma_{I_X}, \sigma_{I_Y} \ll n^{1/2}$.

\end{subassumption}

Similar to Assumption \ref{assump:S2}, Assumption \ref{assump:S1} uses $n^{1/2}$ as a cutoff to describe the scenario where both  the joint and individual structures are   weak. 
In this case, both two types of structures make  marginal contributions  to the overall data variation compared to the noise terms. 
Theorem \ref{thm:power_comparison_S1} implies that the effect of devariation on the power of the RV test is asymptotically negligible.

\begin{theorem}\label{thm:power_comparison_S1} \normalfont
    Under Assumption \ref{assump:S1}, 
    \begin{align}
       &~ \mathbb{P}\big\{\beta_{\alpha}(\bX,\bY)\neq\beta_{\alpha}(\bXd,\bYd)\big\}= O\{\exp(-d_n)\} \to 0\notag 
    \end{align}
as $n\to \infty$, where $d_n = \frac{n}{\sigma_J^2 +\min\{\sigma_{I_X}^2,\sigma_{I_Y}^2\}} $. 
\end{theorem}

\subsection{Scenario 3: dominant joint structure} \label{sec:Scenario3}

\vspace{1em}

\begin{subassumption}\label{assump:S3} \normalfont
Suppose that  Assumption \ref{assump:overall} holds and $\sigma_J \gg n^{1/2}$, 
        $\sigma_{I_X}=\sigma_{I_Y}=0$.
\end{subassumption}

 Under Assumption \ref{assump:S3}, the joint structure constitutes the dominant source of variation compared to white noise. We assume $\sigma_{I_X}=\sigma_{I_Y}=0$ as an extreme adverse scenario where no individualized variation is present. In this setting, devariation is unnecessary and may undesirably remove signal from the joint structure. We establish an asymptotic analysis showing that the RV test still maintains reasonably high asymptotic power.
 The assumption $\sigma_{I_X}=\sigma_{I_Y}=0$ is imposed mainly for notational simplicity, while the same behavior is expected when $\sigma_{I_X}$ and $\sigma_{I_Y}$ are non-zero but sufficiently small. 

\begin{theorem}\label{thm:S3:power_Tr}
\normalfont
Under Assumption \ref{assump:S3}, 
\begin{align}\label{eq:S3:power_Tr}
   \lim_{n\to \infty}   \beta_{\alpha}(\bXd,\bYd)   = 1-\Phi\left( z_{1-\alpha}-r_J\frac{1+\sqrt{2c_Y}+\sqrt{2c_X}+c_Y\sqrt{2c_X}+c_X\sqrt{2c_Y}+2\sqrt{c_Xc_Y}}{\sqrt{c_Xc_Y}\sqrt{2+c_Y(\mu_{E_X,4}-1)+c_X(\mu_{E_Y,4}-1)}}\right).
\end{align}
\end{theorem}

Theorem \ref{thm:S3:power_Tr}  provides the general asymptotic power of the RV test using the devariation-adjusted $\bXd$ and $\bYd$. Based on this result, an asymptotic lower bound for  $\beta_{\alpha}(\bXd, \bYd)$ can be obtained when $\bE_X$ and $\bE_Y$ consist of independent normal entries, in which case
\begin{align}\label{eq:powerlow_bound}
    \lim_{n\to \infty}   \beta_{\alpha}(\bXd,\bYd) \geqslant  1-\Phi\left(z_{1-\alpha}-r_J  \frac{3+4\sqrt{2}}{\sqrt{6}}\right). 
\end{align} 
The proof of \eqref{eq:powerlow_bound} is deferred to the Supplementary Material A.4.5. This lower bound indicates that  the RV test based on devariation-adjusted data still retains high asymptotic power.  As an illustration, when $\alpha = 0.05$ and $r_J = 1$, the bound evaluates to 0.97, indicating that the loss of power due to devariation is negligible relative to the original RV test. 
This result highlights the robustness of the proposed devariation strategy, as it incurs only marginal power loss even in this unfavorable regime. 


\vspace{-10pt}

\paragraph{Theoretical novelty.} 
Because our method intentionally makes use of the traditional `noise' term $\bX-S_{\lambda_X}(\bX)$, 
standard  intuition or tools from denoising large matrices  cannot   directly apply. To address this gap, we  develop a  new analytical framework tailored for quantifying association under the  latent factor model. 
Our theory shows  that the devariation not only achieves  significant power gain under favorable scenarios, but also  maintains reasonably high power in the adverse regimes. 
The  adaptiveness of power is achieved by a non-traditional argument showing  that part of the association signal from the joint structure can be retained in $\bX-S_{\lambda_X}(\bX) $, even though it is conventionally treated as noises in matrix recovery.  Our analysis reverses the standard role of the  residual after soft-thresholding in matrix completion and offers new theoretical insights for multi-view association testing. 

\section{Simulation study}\label{sec:simulation}
We conduct simulation studies to examine the finite-sample behavior of the devariation RV  and to complement the asymptotic power analysis in Section \ref{sec:theory}. Specifically, we focus on (i) comparing the power of devariation RV,  RV, and GEE-based tests under varying relative strengths of joint and individual variations; (ii) examining robustness under model misspecification of the within-view dependence structure; and (iii) investigating the sensitivity of the methods to deviations from Gaussian noise.

\subsection{Gaussian noise}\label{subsec:sim_mis_model}

\subsubsection{Settings}\label{subsubsec:sim_mis_model_settings}
To compare the proposed method with existing methods, data are generated under different within-view dependency structures: low-rank, exchangeable correlation, and autoregressive correlation structures. The baseline setting for all cases is  $n=300$, $p=200$, and $q=250$. The joint structures are sampled as in Equation \eqref{eq:low_rank_model}, with ranks $r_J \in\{5, 10, 15\}$ and noise variances are fixed  at $\tau_X^2=\tau_Y^2=1$.  Then, the individual structures and noises are sampled from different models:
\begin{itemize}[nolistsep]
     \item[1)] \textbf{Low-rank factor model}: The $\sigma_J^2$ for the joint structure is fixed at $0.7$. The individual structure and noise are also sampled from the low-rank model as shown in Equation \eqref{eq:low_rank_model}. We vary the ranks of the individual structures $r_{I_X}=r_{I_Y}\in\{5,10,15\}$ under each $r_J$. To control the strength of individual variation, we vary $\sigma_{I_X} = \sigma_{I_Y} \in\{1, 2, \ldots, 5\}$.  
    \item[2)] \textbf{Exchangeable correlation}: The $\sigma_J^2$ for the joint structure is fixed at 0.5.  We set the individual structure + noise to have an exchangeable correlation structure. In this case, exchangeable correlation is equivalent to a low-rank factor model with  $\bL_{I_X}^\top=(1/\sqrt{p}, \ldots, 1/\sqrt{p})$ and $\bL_{I_Y}^\top=(1/\sqrt{q}, \ldots, 1/\sqrt{q})$ ($r_{I_X}=r_{I_Y}=1$). Thus, we directly sample the data from Equation \eqref{eq:low_rank_model} with these specific loadings for individual structures. We also vary $\sigma_{I_X} = \sigma_{I_Y} \in\{1, 2, \ldots, 5\}$ to control  the strength of individual variation. 
    \item[3)] \textbf{AR(1) {correlation}}: The $\sigma_J^2$ is chosen to be 0.7.  The individual structure plus noise follows an autoregressive correlation structure of order 1 (AR(1)):
    $
        r_{i,k}^{X}=\phi r_{i, k-1}^{X}+\epsilon_{i,k}^{X}
    $
    where $r_{i,k}^{X}$ is the $(i,k)$th entry of $\bS_{I_X}\bL_{I_X}^\top+\bE_X$ and $\epsilon_{i,k}^{X}$ is white noise with $\text{Var}(\epsilon_{i,k}^{X})=\tau_X^2$. Sampling for $\bY$ is analogous. 
    We vary the autocorrelation parameter $\phi\in\lbrace0.35, 0.45, 0.55, 0.75,0.95\rbrace$. We note that {it is an example of model misspecification because} the AR(1) structure cannot be represented as a low-rank factor model.
\end{itemize}
Correspondingly, we consider four competitors:
\begin{enumerate}[nolistsep]
\item \textbf{Gold standard}: $T_{\text{RV}}(\bX\bSigma_{X}^{-1},\bY\bSigma_{Y}^{-1})$ in Equation 
 \eqref{eq:rv_score} {using the true $\bSigma_X$ and $\bSigma_Y$} under $H_0$.
    \item \textbf{RV}: $ T_{\text{RV}}(\bX,\bY)$.
    \item \textbf{GEE-EX}:  $T_{\text{RV}}(\bX\widetilde{\bSigma}_{X}^{-1},\bY\widetilde{\bSigma}_{Y}^{-1})$ where $\widetilde{\bSigma}_{X}^{-1}$ and $ \widetilde{\bSigma}_{Y}^{-1}$ are estimated  assuming exchangeable working correlations.
\item \textbf{GEE-AR(1)}: $T_{\text{RV}}(\bX\widetilde{\bSigma}_{X}^{-1},\bY\widetilde{\bSigma}_{Y}^{-1})$ where $\widetilde{\bSigma}_{X}^{-1}$ and $ \widetilde{\bSigma}_{Y}^{-1}$ are  estimated assuming AR(1) working correlation{s}.

\end{enumerate}
For evaluating the Type I error rate, we repeat each simulation setting 10000 times with $\sigma_J=0$ {and} $\alpha=0.05$. 
To evaluate the power, we repeat each simulation setting 1000 times. 

\subsubsection{Results}

\begin{figure}
    \centering
\includegraphics[width=1\linewidth]{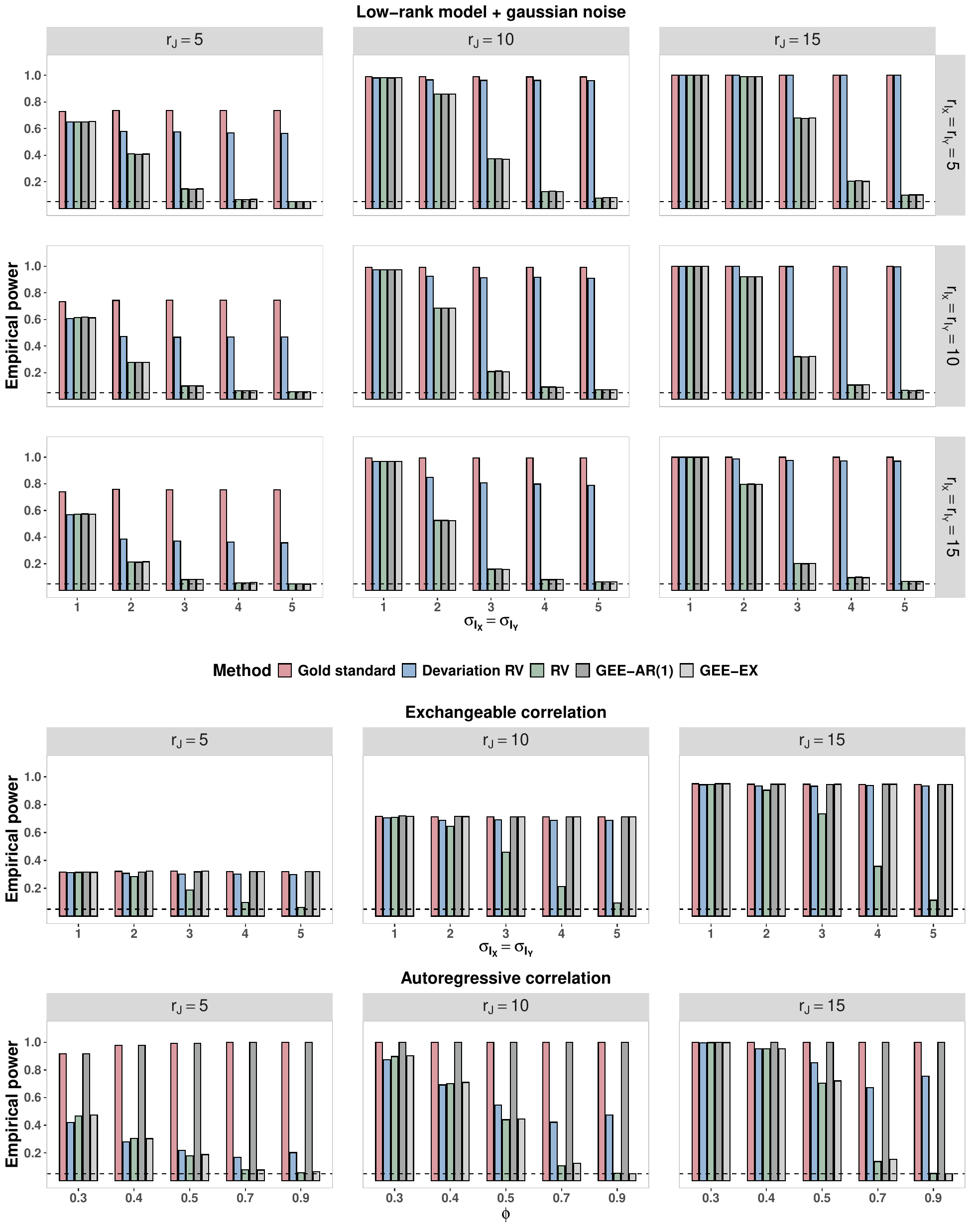}
    \caption{Empirical power for different methods when data are generated from different models.  Significance level $\alpha=0.05$ (dashed line).}
 \label{fig:power_mis_model}
\end{figure}

First, {as expected, }all methods control the Type I error rate accurately at $\alpha=0.05$. Regarding the power performance shown in Figure \ref{fig:power_mis_model},  when the data were generated from the low-rank model, devariation RV outperformed the GEE-based methods, achieving power levels close to that of the gold standard.
As $\sigma_{I_X}$ and $\sigma_{I_Y}$ increased, the difference between devariation RV and other methods {except for the gold standard} became more obvious with stable power for devariation RV but decreasing power for the other methods. Since the low-rank model efficiently captured the strong low-rank individual variation, devariation RV  maintains its power at a high level. However, the exchangeable and AR(1) correlation structures failed to capture the randomly generated low-rank structure, resulting in GEE-AR(1) and GEE-EX having similar performance as RV, and the three methods were heavily affected by the increased individual variation. 

Moving to the setting where within-view dependencies are exchangeable in Figure \ref{fig:power_mis_model}, as expected, the GEE-EX exhibits power nearly identical to that of the gold standard, as the specified covariance matrix effectively captured the exchangeable correlation. The power of the devariation RV  method is also close to the top-performing methods with slightly lower power. This can be explained by the low-rank representation of exchangeable correlation, which allows the proposed method to capture most of the variation. However, GEE-AR(1) achieved significantly lower power than devariation RV  and decreased as the strength of individual variation increased because AR(1) did not efficiently capture the exchangeable correlation. Since AR(1) still captured partial variation from the exchangeable correlation, its power is still higher than the RV  which assumes the independent structure. 

In Scenarios involving AR(1) correlation structures, as shown in the last row of Figure \ref{fig:power_mis_model},  the GEE-AR(1) achieved the best performance {close to the}  gold standard. The devariation RV did not perform well with much lower power than GEE-AR(1) under this model-misspecification scenario  because the AR(1) structure is a full-rank model. However, it performed better than RV and GEE-EX, which failed to capture the AR(1) correlation. We also observed a non-monotonic trend for the power of the devariation RV, which was not the case for RV and GEE-EX. As $\phi$ increased, the power of RV and GEE-EX decreased while devariation RV's power decreased at first, then increased after a certain point. For GEE-EX and RV, since the exchangeable and independent structures are far away from the AR(1) structure, the power of GEE-EX and RV continued to decrease as individual variation increased, which was caused by the increase in  $\phi$. Devariation RV achieved high power when $\phi$ was small, as small $\phi$ implies weak individual variation and a high signal-to-noise ratio. As $\phi$ increased, individual variation increased, leading to decreased power of devariation RV. However, when $\phi$ is large enough, the low-rank model can capture more variation from AR(1) structure, which explains the increased power of the devariation RV when $\phi$ was large.

\subsection{Non-Gaussian noise}

To assess the robustness of devariation RV, {we consider non-}Gaussian noise distributions {for $\bE_X$ and $\bE_Y$} in the low-rank factor model in Equation \eqref{eq:low_rank_model}. Except for the noise type, we keep the other settings consistent with those in Section \ref{subsubsec:sim_mis_model_settings}. Following \cite{choi2017selecting}'s simulation designs, we examined (i) heavy-tailed and (ii) right-skewed noises. For heavy-tailed noise, we sample from $\sqrt{\frac{3}{5}}t_5$ where $t_5$ denotes the $t\text{-distribution}$ with degrees of freedom 5. The right-skewed noise is sampled from $\sqrt{\frac{3}{10}}t_5+\sqrt{\frac{1}{2}}(\text{{Exp}}(1)-1)$ where $\text{Exp}(1)$ denotes the exponential distribution with {rate} 1. 

\begin{figure}
    \centering
    \includegraphics[width=1\linewidth]{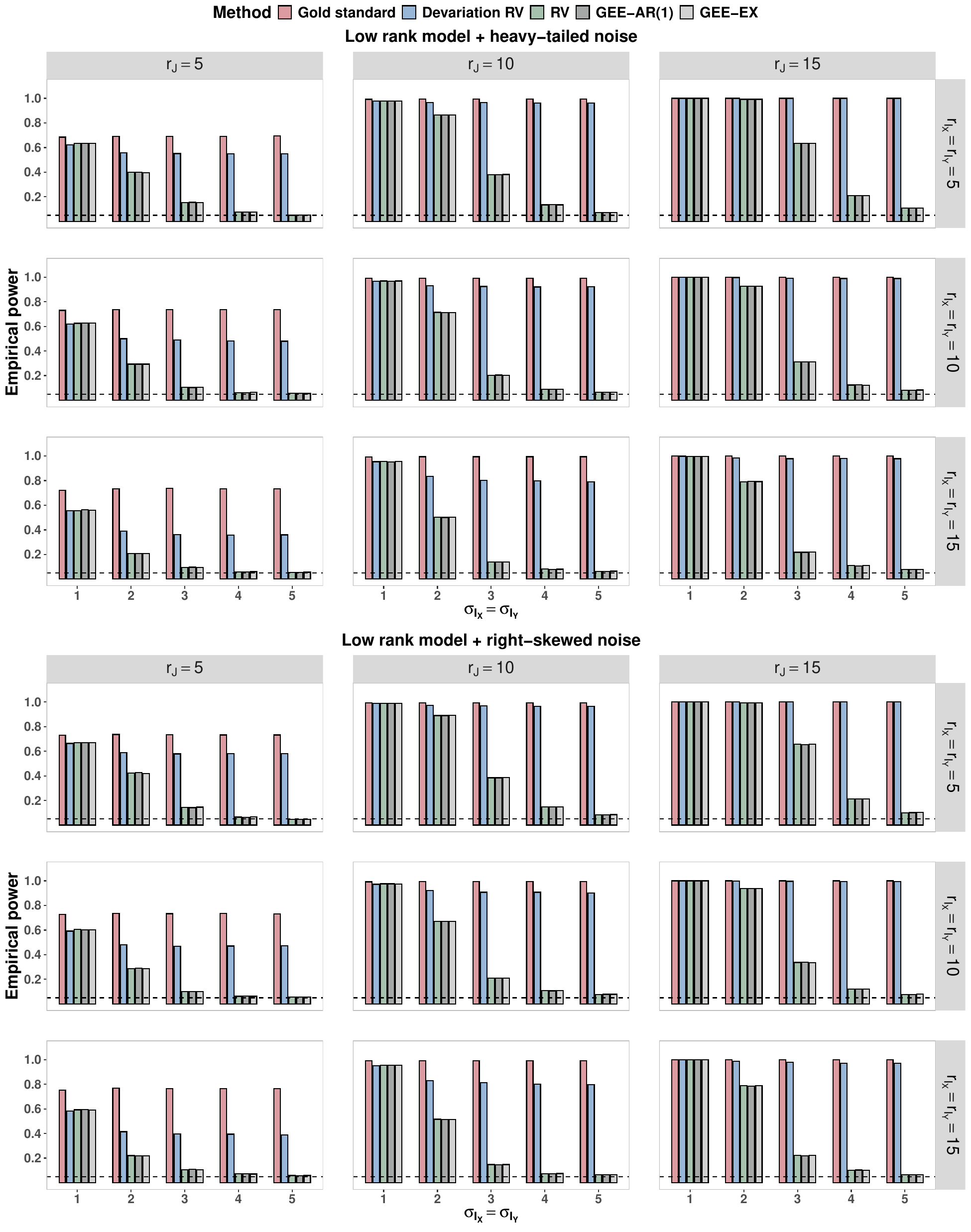}
    \caption{Empirical power for different methods when data are generated from low rank model with different types of noise.  Significance level $\alpha=0.05$ (dashed line).}
    \label{fig:power_noise}
\end{figure}

Figure \ref{fig:power_noise} shows the empirical power of each method under different noises. Comparing them to the results for the low-rank model with Gaussian noise in Figure \ref{fig:power_mis_model}, we observe that the noise types only marginally affect the power performance when the low-rankness holds. Overall, devariation RV still significantly outperformed GEE-based methods and RV, demonstrating its robustness in the presence of non-Gaussian noise{,  which aligns with the theoretical results in Section \ref{sec:theory}}.

\section{Real data analysis}\label{sec:real_data}
We conducted data-driven simulations to compare the performance of the proposed method and the RV test. We used brain IDPs from three views: fMRI, sMRI, and dMRI obtained from the UK Biobank study (\url{https://www.ukbiobank.ac.uk}). The study subjects were de-identified, and UK Biobank obtained ethical approval from the relevant research ethics committee, and all participants provided informed consent. Ethical approval for our analysis was reviewed and granted by the Office of Research Ethics, University of Toronto (protocol number 44674). There are 39587 subjects, with 210 features in fMRI, 339 features in sMRI, and 432 features in dMRI. The descriptions of the features are shown in Figure \ref{fig:corrplot_full}. We  regressed out age, biological sex, and study sites to prevent spurious associations. 

\subsection{Type I error}
To evaluate the Type I error rate, we randomly sampled non-overlapping sets of subjects from the three views with varying sample sizes ($n=50, 100, 150, 200$) and repeated the procedure 10000 times to obtain empirical Type I error rates across three pairwise combinations (sMRI $\leftrightarrow$ dMRI, sMRI $\leftrightarrow$ fMRI, dMRI $\leftrightarrow$ fMRI). As expected, devariation RV and RV accurately controlled the Type I error rate at $\alpha=0.001$ and $0.01$.

\subsection{Power}
To assess the power of the devariation RV in real data, we randomly sampled $n$ subjects (with $n=50, 100, 150, 200$) and obtained their corresponding observations from the three data views. This process was repeated 2000 times to obtain the empirical power of the devariation RV and its competitors (RV, GEE-AR(1), and GEE-EX) across the three pairs of data views.

\begin{figure}[h!]
    \centering
    \includegraphics[scale=0.6]{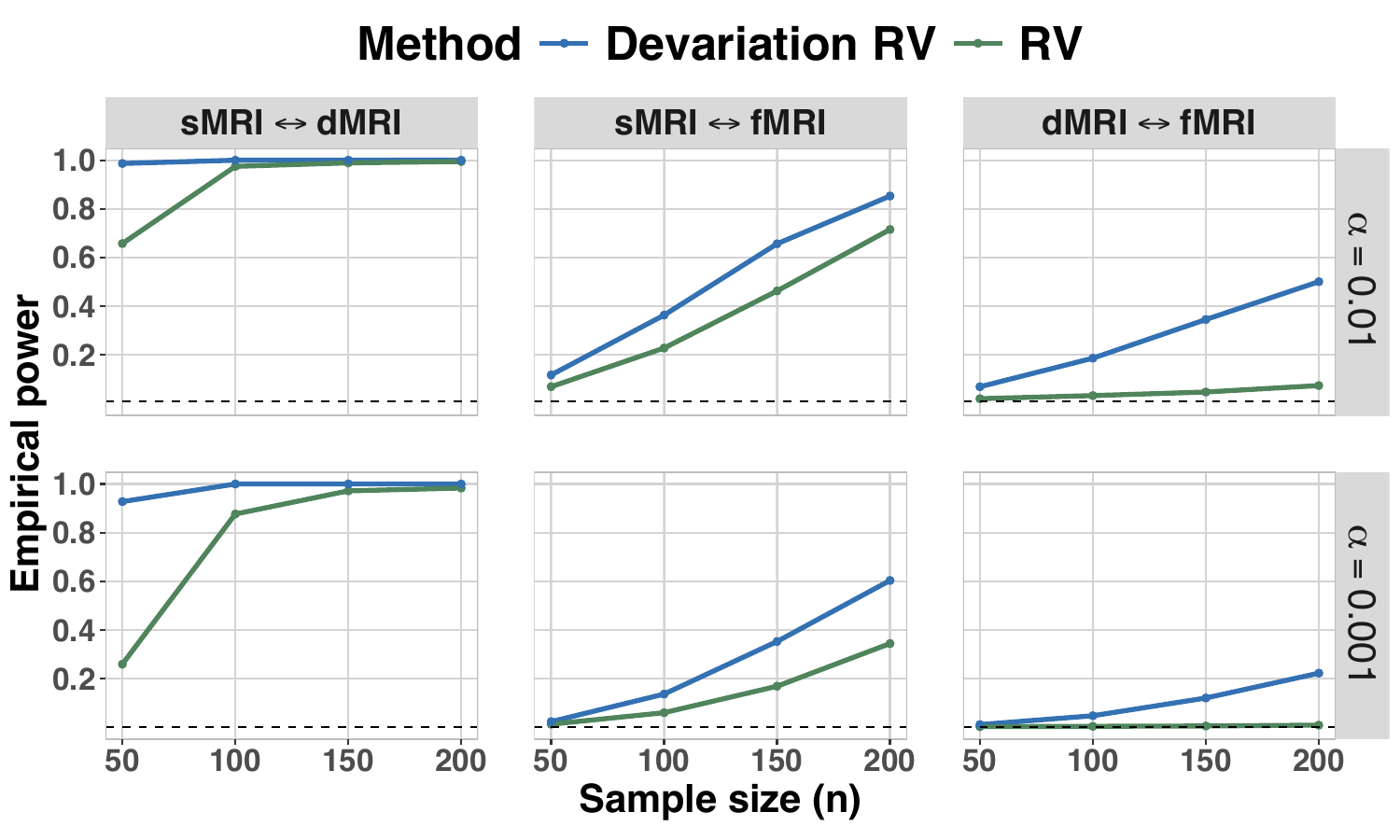}
    \caption{Empirical power results of standard RV and devariation RV under different significance levels ($\alpha=0.01, 0.001$) across three pairs of data views.}
    \label{fig:power_drv}
\end{figure}

As shown in Figure \ref{fig:power_drv}, devariation RV outperformed RV across all settings. Since the performance of GEE-based methods is almost the same as the  RV test, we did not include the power results of GEE-based methods and instead only compared the power of devariation RV and the RV tests.  Both methods showed the strongest evidence of the association between sMRI and dMRI, but the weakest evidence for the association between dMRI and fMRI, which is consistent with the overall largest Pearson correlations between sMRI and dMRI and the smallest Pearson correlations between dMRI and fMRI observed in Figure \ref{fig:corrplot_full}. 
Devariation RV's power was significantly higher than the RV's power when testing the association between dMRI and fMRI, aligning with our theoretical expectations for Scenario 2 in Section \ref{sec:Scenario2}. Specifically, from the Pearson correlation matrix shown in Figure \ref{fig:corrplot_full}, the within-view dependencies specific to dMRI and fMRI are much stronger than the across-view dependencies between dMRI and fMRI. As anticipated, devariation, by capturing the strong within-view dependencies, led to an improvement in power most notably in this case. Although their Pearson correlations show that the within-view dependencies are non-homogeneous across the three views, devariation RV consistently improved power, showing its robustness across different datasets.

\section{Discussion}\label{sec:discussion}
In this paper, we propose devariation as a novel testing procedure for high-dimensional multi-view data, where, when applied to the RV test, the singular values of each data view are soft-thresholded at certain levels before applying the RV test. The method is practically appealing because it avoids the need to specify a working covariance structure or choose a kernel, both of which are often not straightforward in biomedical data, where dependence patterns are often complex or nontrivial. Instead, our method uses random matrix theory to determine the threshold level, yielding a simple and interpretable method that is easy to implement, computationally efficient, and compatible with standard inference procedures. 

A critical part of our work is the theoretical analysis we developed under the JIVE model, which is a promising and realistic model for biomedical data in characterizing variation into several interpretable components. Although JIVE and related low-rank methods have been widely used for exploratory decomposition and integrative analysis, their role in formal hypothesis testing has remained largely unexplored. To our knowledge, this is the first study to explicitly characterize the asymptotic power of an RV-based association test under a JIVE-type latent factor model and to show how singular value soft-thresholding affects power in high-dimensional settings. Our results show that devariation RV can substantially improve power when individual variation dominates and obscures cross-view associations, while remaining robust in less favorable settings where individual variation is minimal.


In this paper, we focused primarily on applying devariation to the RV test as it allows for an explicit parametrization under the JIVE model followed by the score-based variance component testing that takes a quadratic form as a test statistic, making the asymptotic power analysis feasible. Although we have applied devariation-adjusted data to dCor, HSIC, and other variants (e.g., \citet{xu2017adaptive}) and observed a noticeable power gain in both simulation studies and real data analyses (not shown in this paper), its model-based interpretation or underlying mechanism in terms of power is not too straightforward. Exploring theoretical properties of devariation for other association tests would be an interesting research direction  to support its generality.

There are several potential directions for extending the devariation approach. The JIVE model assumes the low-rankness of the joint variation, it is expected that the RV coefficient, which takes all singular values of $\bX^\top\bY$, could be underpowered. Association tests that directly utilize this low-rank assumption would be worth further investigation (e.g., tests based on the Ky-Fan$(k)$ norm \citep{veitch2023rank}). In addition, although the tuning parameters $\lambda_X$ and $\lambda_Y$, selected based on random matrix theory, provide asymptotic power guarantees under certain conditions, they are not guaranteed to be optimal for maximizing power. Developing adaptive or data-driven tuning strategies could further improve performance.
Also, as shown in our simulations, devariation RV's performance was significantly worse than GEE-AR(1) when the within-view dependence exhibits a full-rank structure, such as AR(1) correlations. Future research could explore ways to integrate more flexible dependency models into the devariation framework to expand its generality. 

\section*{Software}
An implementation of the devariation RV in the form of an R package can be found at \url{https://github.com/RuyiPan/Devariation}.

\section*{Supplementary material}
Proofs of theorems in Section \ref{sec:theory} can be found in the supplementary material.

\section*{Disclosure statement}

The authors report there are no competing interests to declare. 

\section*{Funding}
This work was supported by the McLaughlin Centre, University of Toronto (Accelerator grant, PI: JYP). YH was partially supported by the  Wisconsin Alumni Research Foundation and National Science Foundation (DMS-2515523). 
JYP was partially supported by the
Natural Sciences and Engineering Research Council of Canada (RGPIN-2022-04831) the University of Toronto’s Data Sciences Institute, and the Connaught Fund. The computing resources were enabled in part by support provided by University of Toronto and the Digital Research Alliance of Canada.

\section*{Data availability statement}

The data that support the findings of this study were obtained from UK Biobank under approved application 64875. Restrictions apply to the availability of these data, but they are available to eligible researchers upon application to and approval by UK Biobank.

\bibliographystyle{apalike}
{\singlespacing
\bibliography{ref}

@article{robert1976unifying,
  title={{A unifying tool for linear multivariate statistical methods: the RV-coefficient}},
  author={Robert, Paul and Escoufier, Yves},
  journal={Journal of the Royal Statistical Society Series C: Applied Statistics},
  volume={25},
  number={3},
  pages={257--265},
  year={1976},
  publisher={Oxford University Press}
}

@article{donoho1995noising,
  title={De-noising by soft-thresholding},
  author={Donoho, David L},
  journal={IEEE Transactions on Information Theory},
  volume={41},
  number={3},
  pages={613--627},
  year={1995},
  publisher={IEEE}
}

@article{cai2010singular,
  title={A singular value thresholding algorithm for matrix completion},
  author={Cai, Jian-Feng and Cand{\`e}s, Emmanuel J and Shen, Zuowei},
  journal={SIAM Journal on Optimization},
  volume={20},
  number={4},
  pages={1956--1982},
  year={2010},
  publisher={SIAM}
}

@article{bartlett1941statistical,
  title={The statistical significance of canonical correlations},
  author={Bartlett, Maurice S},
  journal={Biometrika},
  volume={32},
  number={1},
  pages={29--37},
  year={1941},
  publisher={JSTOR}
}

@article{josse2008testing,
  title={{Testing the significance of the RV coefficient}},
  author={Josse, Julie and Pag{\`e}s, J{\'e}rome and Husson, Fran{\c{c}}ois},
  journal={Computational Statistics \& Data Analysis},
  volume={53},
  number={1},
  pages={82--91},
  year={2008},
  publisher={Elsevier}
}

@article{heo1998permutation,
  title={{A permutation test of association between configurations by means of the RV coefficient}},
  author={Heo, Moonseong and Ruben Gabriel, K},
  journal={Communications in Statistics-Simulation and Computation},
  volume={27},
  number={3},
  pages={843--856},
  year={1998},
  publisher={Taylor \& Francis}
}

@article{wu2011rare,
  title={Rare-variant association testing for sequencing data with the sequence kernel association test},
  author={Wu, Michael C and Lee, Seunggeun and Cai, Tianxi and Li, Yun and Boehnke, Michael and Lin, Xihong},
  journal={The American Journal of Human Genetics},
  volume={89},
  number={1},
  pages={82--93},
  year={2011},
  publisher={Elsevier}
}

@article{szekely2007measuring,
  title={Measuring and testing dependence by correlation of distances},
  author={Sz{\'e}kely, G{\'a}bor J. and Rizzo, Maria L. and Bakirov, Nail K.},
  journal={The Annals of Statistics},
  volume={35},
  number={6},
  pages={2769--2794},
  year={2007},
  publisher={Institute of Mathematical Statistics}
}

@inproceedings{gretton2005measuring,
  title={{Measuring statistical dependence with Hilbert-Schmidt norms}},
  author={Gretton, Arthur and Bousquet, Olivier and Smola, Alex and Sch{\"o}lkopf, Bernhard},
  booktitle={International Conference on Algorithmic Learning Theory},
  pages={63--77},
  year={2005},
  organization={Springer}
}

@article{berrett2019nonparametric,
  title={Nonparametric independence testing via mutual information},
  author={Berrett, Thomas B and Samworth, Richard J},
  journal={Biometrika},
  volume={106},
  number={3},
  pages={547--566},
  year={2019},
  publisher={Oxford University Press}
}

@article{heller2013consistent,
  title={A consistent multivariate test of association based on ranks of distances},
  author={Heller, Ruth and Heller, Yair and Gorfine, Malka},
  journal={Biometrika},
  volume={100},
  number={2},
  pages={503--510},
  year={2013},
  publisher={Oxford University Press}
}

@article{wang2013gee,
  title={{GEE-based SNP set association test for continuous and discrete traits in family-based association studies}},
  author={Wang, Xuefeng and Lee, Seunggeun and Zhu, Xiaofeng and Redline, Susan and Lin, Xihong},
  journal={Genetic Epidemiology},
  volume={37},
  number={8},
  pages={778--786},
  year={2013},
  publisher={Wiley Online Library}
}

@article{dutta2019multi,
  title={{Multi-SKAT: General framework to test for rare-variant association with multiple phenotypes}},
  author={Dutta, Diptavo and Scott, Laura and Boehnke, Michael and Lee, Seunggeun},
  journal={Genetic Epidemiology},
  volume={43},
  number={1},
  pages={4--23},
  year={2019},
  publisher={Wiley Online Library}
}

@article{xu2017adaptive,
  title={Adaptive testing for association between two random vectors in moderate to high dimensions},
  author={Xu, Zhiyuan and Xu, Gongjun and Pan, Wei and Alzheimer's Disease Neuroimaging Initiative},
  journal={Genetic Epidemiology},
  volume={41},
  number={7},
  pages={599--609},
  year={2017},
  publisher={Wiley Online Library}
}

@article{zhang2014testing,
  title={Testing for association with multiple traits in generalized estimation equations, with application to neuroimaging data},
  author={Zhang, Yiwei and Xu, Zhiyuan and Shen, Xiaotong and Pan, Wei and Alzheimer's Disease Neuroimaging Initiative and others},
  journal={Nueroimage},
  volume={96},
  pages={309--325},
  year={2014},
  publisher={Elsevier}
}

@article{wang2003working,
  title={Working correlation structure misspecification, estimation and covariate design: implications for generalised estimating equations performance},
  author={Wang, You-Gan and Carey, Vincent},
  journal={Biometrika},
  volume={90},
  number={1},
  pages={29--41},
  year={2003},
  publisher={Oxford University Press}
}

@article{zapala2006multivariate,
  title={Multivariate regression analysis of distance matrices for testing associations between gene expression patterns and related variables},
  author={Zapala, Matthew A and Schork, Nicholas J},
  journal={Proceedings of the National Academy of Sciences},
  volume={103},
  number={51},
  pages={19430--19435},
  year={2006},
  publisher={National Acad Sciences}
}

@article{han2010powerful,
  title={Powerful multi-marker association tests: unifying genomic distance-based regression and logistic regression},
  author={Han, Fang and Pan, Wei},
  journal={Genetic Epidemiology},
  volume={34},
  number={7},
  pages={680--688},
  year={2010},
  publisher={Wiley Online Library}
}

@article{shi2023distance,
  title={Distance-based regression analysis for measuring associations},
  author={Shi, Yuke and Zhang, Wei and Liu, Aiyi and Li, Qizhai},
  journal={Journal of Systems Science and Complexity},
  volume={36},
  number={1},
  pages={393--411},
  year={2023},
  publisher={Springer}
}

@article{zhan2017fast,
  title={A fast small-sample kernel independence test for microbiome community-level association analysis},
  author={Zhan, Xiang and Plantinga, Anna and Zhao, Ni and Wu, Michael C},
  journal={Biometrics},
  volume={73},
  number={4},
  pages={1453--1463},
  year={2017},
  publisher={Oxford University Press}
}

@article{minas2013distance,
  title={A distance-based test of association between paired heterogeneous genomic data},
  author={Minas, Christopher and Curry, Edward and Montana, Giovanni},
  journal={Bioinformatics},
  volume={29},
  number={20},
  pages={2555--2563},
  year={2013},
  publisher={Oxford University Press}
}

@article{broadaway2016statistical,
  title={A statistical approach for testing cross-phenotype effects of rare variants},
  author={Broadaway, K Alaine and Cutler, David J and Duncan, Richard and Moore, Jacob L and Ware, Erin B and Jhun, Min A and Bielak, Lawrence F and Zhao, Wei and Smith, Jennifer A and Peyser, Patricia A and others},
  journal={The American Journal of Human Genetics},
  volume={98},
  number={3},
  pages={525--540},
  year={2016},
  publisher={Elsevier}
}

@article{park2022clean,
  title={{CLEAN: Leveraging spatial autocorrelation in neuroimaging data in clusterwise inference}},
  author={Park, Jun Young and Fiecas, Mark},
  journal={Neuroimage},
  volume={255},
  pages={119192},
  year={2022},
  publisher={Elsevier}
}

@article{mazumder2010spectral,
  title={Spectral regularization algorithms for learning large incomplete matrices},
  author={Mazumder, Rahul and Hastie, Trevor and Tibshirani, Robert},
  journal={The Journal of Machine Learning Research},
  volume={11},
  pages={2287--2322},
  year={2010},
  publisher={JMLR. org}
}

@article{shabalin2013reconstruction,
  title={Reconstruction of a low-rank matrix in the presence of Gaussian noise},
  author={Shabalin, Andrey A and Nobel, Andrew B},
  journal={Journal of Multivariate Analysis},
  volume={118},
  pages={67--76},
  year={2013},
  publisher={Elsevier}
}

@article{gavish2017optimal,
  title={Optimal shrinkage of singular values},
  author={Gavish, Matan and Donoho, David L},
  journal={IEEE Transactions on Information Theory},
  volume={63},
  number={4},
  pages={2137--2152},
  year={2017},
  publisher={IEEE}
}

@inproceedings{rudelson2010non,
  title={Non-asymptotic theory of random matrices: extreme singular values},
  author={Rudelson, Mark and Vershynin, Roman},
  booktitle={Proceedings of the International Congress of Mathematicians 2010 (ICM 2010) (In 4 Volumes) Vol. I: Plenary Lectures and Ceremonies Vols. II--IV: Invited Lectures},
  pages={1576--1602},
  year={2010},
  organization={World Scientific}
}

@article{lock2013joint,
  title={{Joint and individual variation explained (JIVE) for integrated analysis of multiple data types}},
  author={Lock, Eric F and Hoadley, Katherine A and Marron, James Stephen and Nobel, Andrew B},
  journal={The Annals of Applied Statistics},
  volume={7},
  number={1},
  pages={523},
  year={2013},
  publisher={NIH Public Access}
}

@article{feng2018angle,
  title={Angle-based joint and individual variation explained},
  author={Feng, Qing and Jiang, Meilei and Hannig, Jan and Marron, JS},
  journal={Journal of Multivariate Analysis},
  volume={166},
  pages={241--265},
  year={2018},
  publisher={Elsevier}
}

@article{shu2020d,
  title={{D-CCA: A decomposition-based canonical correlation analysis for high-dimensional datasets}},
  author={Shu, Hai and Wang, Xiao and Zhu, Hongtu},
  journal={Journal of the American Statistical Association},
  volume={115},
  number={529},
  pages={292--306},
  year={2020},
  publisher={Taylor \& Francis}
}

@article{park2020integrative,
  title={Integrative factorization of bidimensionally linked matrices},
  author={Park, Jun Young and Lock, Eric F},
  journal={Biometrics},
  volume={76},
  number={1},
  pages={61--74},
  year={2020},
  publisher={Wiley Online Library}
}

@article{choi2017selecting,
  title={Selecting the number of principal components: Estimation of the true rank of a noisy matrix},
  author={Choi, Yunjin and Taylor, Jonathan and Tibshirani, Robert},
  journal={The Annals of Statistics},
  pages={2590--2617},
  year={2017},
  publisher={JSTOR}
}

@article{van2013wu,
  title={{The WU-Minn Human Connectome Project: an overview}},
  author={Van Essen, David C and Smith, Stephen M and Barch, Deanna M and Behrens, Timothy EJ and Yacoub, Essa and Ugurbil, Kamil and Wu-Minn HCP Consortium and others},
  journal={Neuroimage},
  volume={80},
  pages={62--79},
  year={2013},
  publisher={Elsevier}
}

@article{veitch2023rank,
  title={Rank-adaptive covariance testing with applications to genomics and neuroimaging},
  author={Veitch, David and He, Yinqiu and Park, Jun Young},
  journal={Biometrics (in press)},
  year={2026}
}

@article{weinstein2022spatially,
  title={Spatially-enhanced clusterwise inference for testing and localizing intermodal correspondence},
  author={Weinstein, Sarah M and Vandekar, Simon N and Baller, Erica B and Tu, Danni and Adebimpe, Azeez and Tapera, Tinashe M and Gur, Ruben C and Gur, Raquel E and Detre, John A and Raznahan, Armin and others},
  journal={Neuroimage},
  volume={264},
  pages={119712},
  year={2022},
  publisher={Elsevier}
}

@article{zhao2023heart,
  title={Heart-brain connections: Phenotypic and genetic insights from magnetic resonance images},
  author={Zhao, Bingxin and Li, Tengfei and Fan, Zirui and Yang, Yue and Shu, Juan and Yang, Xiaochen and Wang, Xifeng and Luo, Tianyou and Tang, Jiarui and Xiong, Di and others},
  journal={Science},
  volume={380},
  number={6648},
  pages={abn6598},
  year={2023},
  publisher={American Association for the Advancement of Science}
}

@article{lock2022bidimensional,
  title={Bidimensional linked matrix factorization for pan-omics pan-cancer analysis},
  author={Lock, Eric F and Park, Jun Young and Hoadley, Katherine A},
  journal={The Annals of Applied Statistics},
  volume={16},
  number={1},
  pages={193},
  year={2022}
}

@article{lin1997variance,
  title={Variance component testing in generalised linear models with random effects},
  author={Lin, Xihong},
  journal={Biometrika},
  volume={84},
  number={2},
  pages={309--326},
  year={1997},
  publisher={Oxford University Press}
}

@article{pan2024spatial,
  title={Spatial-extent inference for testing variance components in reliability and heritability studies},
  author={Pan, Ruyi and Dickie, Erin W and Hawco, Colin and Reid, Nancy and Voineskos, Aristotle N and Park, Jun Young},
  journal={Imaging Neuroscience},
  volume={2},
  pages={imag--2},
  year={2024},
  publisher={MIT Press One Broadway, 12th Floor, Cambridge, Massachusetts 02142, USA~…}
}
}
\appendix


\counterwithin*{equation}{section}
\renewcommand\theequation{\thesection.\arabic{equation}}

\renewcommand{\theproposition}{\Alph{section}.\arabic{subsection}.\arabic{subsubsection}.\arabic{proposition}}

\newpage

\end{document}